\documentclass[pre,twocolumn,superscriptaddress,showpacs,%
preprintnumbers,amsmath,amssymb]{revtex4}
\usepackage{amsmath}
\usepackage{graphicx}
\usepackage{epsfig}
\usepackage{dcolumn}

\begin{document}
\title{Oil displacement through a porous medium with a temperature gradient}

\author{Cl\'audio L. N. Oliveira}
\affiliation{Departamento de F\'isica, Universidade Federal do
Cear\'a, Caixa Postal 6030, Campus do Pici, 60451-970 Fortaleza,
Cear\'a, Brazil} \affiliation{Computational Physics, IfB, ETH
Z\"urich, H\"onggerberg, 8093 Z\"urich, Switzerland}
\author{Jos\'e S. Andrade Jr.}
\affiliation{Departamento de F\'isica, Universidade Federal do
Cear\'a, Caixa Postal 6030, Campus do Pici, 60451-970 Fortaleza,
Cear\'a, Brazil} \affiliation{Computational Physics, IfB, ETH
Z\"urich, H\"onggerberg, 8093 Z\"urich, Switzerland}
\author{Hans J. Herrmann}
\affiliation{Departamento de F\'isica, Universidade Federal do
Cear\'a, Caixa Postal 6030, Campus do Pici, 60451-970 Fortaleza,
Cear\'a, Brazil} \affiliation{Computational Physics, IfB, ETH
Z\"urich, H\"onggerberg, 8093 Z\"urich, Switzerland}


\begin{abstract}
  We investigate the effect of a temperature gradient on oil recovery
  in a two-dimensional pore-network model. The oil viscosity depends
  on temperature as, $\mu_o=exp(B/T)$, where $B$ is a
  physico-chemical parameter depending on the type of oil, and $T$ is
  the temperature. A temperature gradient is applied across the medium
  in the flow direction. Initially, the porous medium is saturated
  with oil and, then, another fluid is injected. We have considered
  two cases representing different injection strategies. In the first
  case, the invading fluid viscosity is constant (finite viscosity
  ratio) while in the second one, the invading fluid is inviscid
  (infinite viscosity ratio). Our results show that, for the case of
  finite viscosity ratio, recovery increases with $\Delta T$
  independently on strength or sign of the gradient. For an infinite
  viscosity ratio, a positive temperature gradient is necessary to
  enhance recovery. Moreover, we show that, for $\Delta T>0$, the
  percentage of oil recovery generally decreases (increases) with $B$
  for a finite (infinite) viscosity ratio. Finally, we also extend our
  results for infinite viscosity ratio to a three-dimensional porous
  media geometry.
\end{abstract}


\maketitle

\section{Introduction}
\label{sec:introduction}

Thermal recovery processes have been used by petroleum companies as a
strategic method to improve oil production from reservoirs. These
processes consist, basically, in decreasing the oil viscosity
(increasing the pressure) by increasing the temperature of the
reservoir using a heat source \cite{Prats}. In practice, this can be
done by injecting a hot fluid (steam or water) into the reservoir. A
method often called Steam or Hot-Water Injection has been mostly used
by companies exploiting heavy oil reservoirs \cite{Craft,Moore}.

The temperature dependence of the oil viscosity is governed by
physico-chemical parameters which can allow a reduction of oil
viscosity by several orders of magnitude with only a modest increase
of temperature \cite{Prats}.
%
%
In general, oil properties in the operational conditions of a
reservoir field are very difficult to predict and can be very
different for each type of oil or oil mixture.  Therefore, their
direct measurement is highly desirable for a better understanding of
this phenomenon, and to improve the efficiency of the recovery.

Several approaches to model and simulate oil recovery have been
utilized in the past. Some of them make use of the macroscopic
description of conservation laws in a porous medium, simulating a
whole reservoir, including injectors and producers wells
\cite{Turta,Kuhlman}. Other authors use conservation laws under a more
microscopic approach
\cite{Andrade,Ferer,Ferer1,Andrade2,Andrade3,Auradou,Andrade4,Andrade5,Sahimi,Andrade6,Andrade7,Lenormand,Andrade8},
where a portion of the porous medium can be represented by tubes
connected to one another. The fluid flow in each tube is easily
computed by the Hagen-Pouseille equation. For instance, Lu {\it et
  al.}  \cite{Lu,Lu1,Lu2} have used such type of modeling approach to
simulate oil burning by air injection. This represents another
thermal recovery method, where the heat source is the burning oil
itself. In their work, they have studied the penetration of the
burning front into a solid oil phase.
%
%
In the present work, we study the displacement in a porous medium of
an oil with temperature-dependent viscosity being pushed at
microscopic scale by other fluid.
%
%
We adopt a simple two-dimensional network model \cite{Aker,Aker2}
previously developed to simulate two-phase flow with arbitrary
viscosity ratio. Despite its simplicity, the model is capable to
reproduce a large variety of experimental results \cite{Aker}.  In
order to adapt this model to our purpose, we implement a temperature
gradient in the injection direction and assume that the oil
viscosity has an exponential dependence of on the inverse of
temperature \cite{Prats}.
%
%
We studied two different cases according to the viscosity ratio. In
the first one, the invading fluid viscosity is constant (finite
viscosity ratio) and in the second one, the invading fluid is inviscid
(infinite viscosity ratio). The aim of this work is then to
investigate the influence of a temperature gradient on the efficiency
of oil recovery under these different conditions.

\section{Model Formulation}
\label{sec:model}

The disordered porous medium is represented by links and nodes. The
links represent pieces of rock of equal length $\ell$ and
cross-section area $a$, and to each one we assign a permeability $k$,
which is chosen randomly according to a uniform distribution in the
interval $[10^{-5},1]$.  This randomness in the permeability
represents the disorder of the porous medium. The nodes where four
links meet are assumed to have no volume. The links are placed on a
square lattice tilted by 45 degrees which assures that all links are
geometrically equivalent with regard to the average flow, i.e., the
links are neither parallel nor perpendicular to the flow direction.
Initially, the porous medium is fully saturated with oil and periodic
boundary conditions are applied at the top and bottom of the system.
The penetration process starts with an invading fluid being injected
at constant flow rate through the left boundary of the system.

The volumetric flow rate in a link connecting neighbor nodes $i$ and
$j$ is given by the Darcy's law,
\begin{equation}
q_{ij} = -\frac{k_{ij}}{\mu_{ef}}\frac{(p_j-p_i)}{\ell},
\label{eq:darcy}
\end{equation}
where $p_i$ is the pressure at node $i$ and $k_{ij}$ is the
permeability of the link. The effective viscosity $\mu_{ef}$ for a
given volume link is calculated according to a linear mixing rule,
namely, $\mu_{ef}=S_o \mu_o+S_I \mu_I$, where $S_o$ and $S_I$
are the saturations of the oil and the invading fluid, respectively,
and $\mu_o$ and $\mu_I$ are their corresponding viscosities. For each
species, the saturation is calculated at a link as the volume fraction
of the corresponding phase. We assume that a link is immediately
accessible when touched by the invading fluid at one end,
%
%
neglecting wetting or drying effects, the pinning of the interface due
to impurities, and finite contact angles or surface tension at the
pore level. The capillary forces in our model are implicit in the
permeability of the link and depend thus on the scale of this link.
Therefore we cannot calculate an explicit flow rate in physical units.

Mass conservation at each node of the lattice leads to the following
set of coupled linear algebraic equations:
\begin{equation}
\sum_{j}q_{ij} = 0, \,\,\, for \,\,\, i=1,2,...,N,
\label{eq:conservation}
\end{equation}
where $N=L^2$ is the total number of nodes, $L$ is the linear size of
the lattice in the $x$-direction, and the summation $j$ runs over the
nearest neighbor nodes of node $i$. These equations are solved to
obtain the node pressures at each time step. In order to simulate the
dynamics of viscous invasion, we neglect the effects of fingers in
each link containing both phases and consider an abrupt saturation
profile along its axial direction. Thus we define an interface inside
the link separating the oil and the invading fluid that is an
approximation and should not be confused with a meniscus in a pore.
Then, we allow those interfaces to displace by a length, $\Delta
x_{ij} = q_{ij}\Delta t_{min}/a$, where $\Delta t_{min}$ is the
minimum time, among all links containing both phases, necessary for
the invading fluid to reach the end of a link. When an interface
reaches the end of a link, reaching a point-like node, it is
instantaneously transferred to those neighbour links whose pressure
differences allow for oil displacement. To avoid multiple interfaces
in a single link the following rule is adopted.  When a third
interface appears in a link, these three interfaces are reduced to a
single one by merging bubbles of the same phase, so that the phase
volume in each link is conserved. The unphysical jumps on the pressure
resulting from this reorganization scheme represent only negligible
perturbations due to the small size of the corresponding bubbles
\cite{Aker}. This procedure is executed at each time step until
breakthrough happens, i.e., the invading fluid just reaches the other
end of the system .
\begin{figure}
\begin{center}
\includegraphics*[scale=0.26]{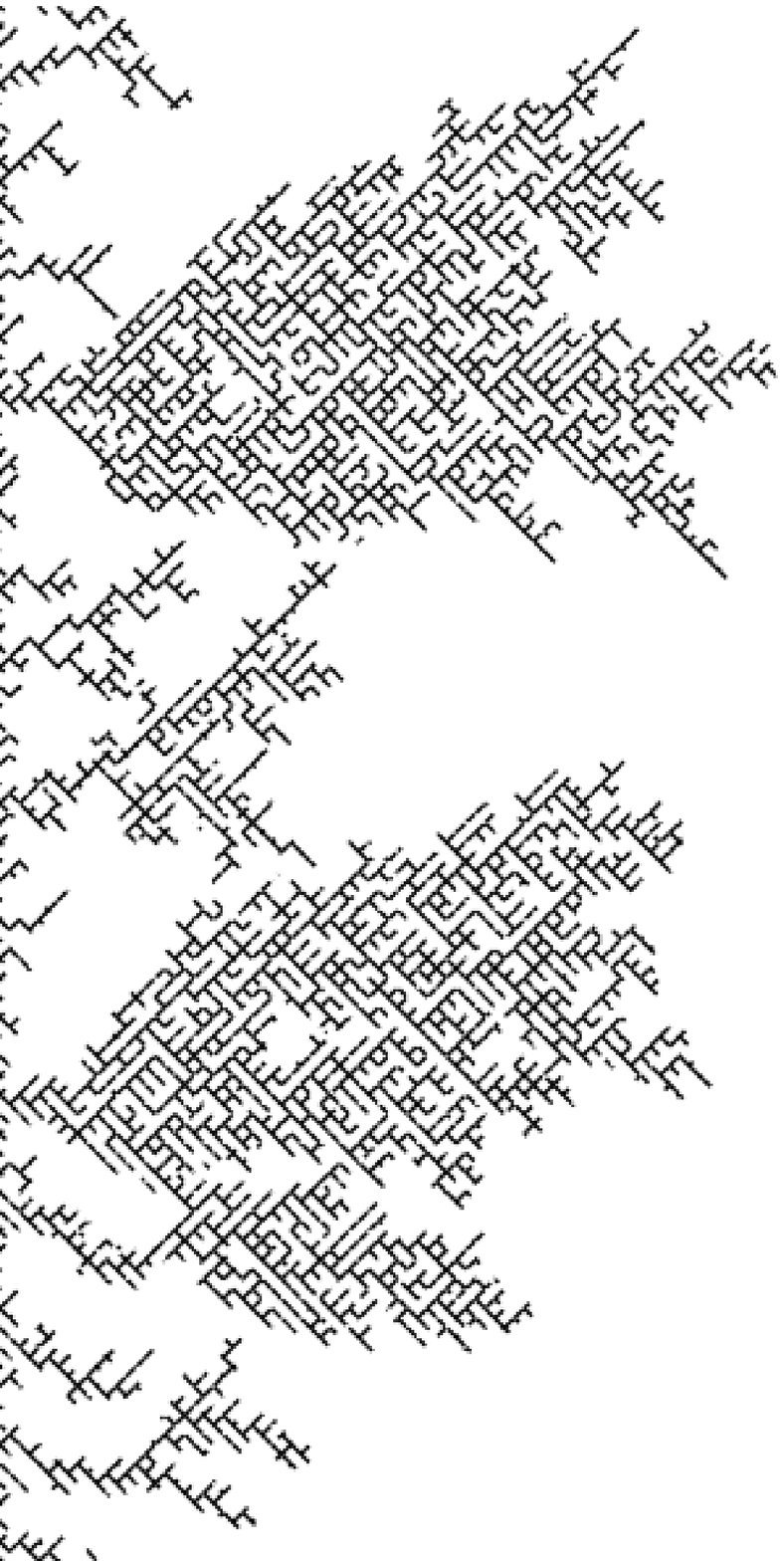}
\includegraphics*[scale=0.26]{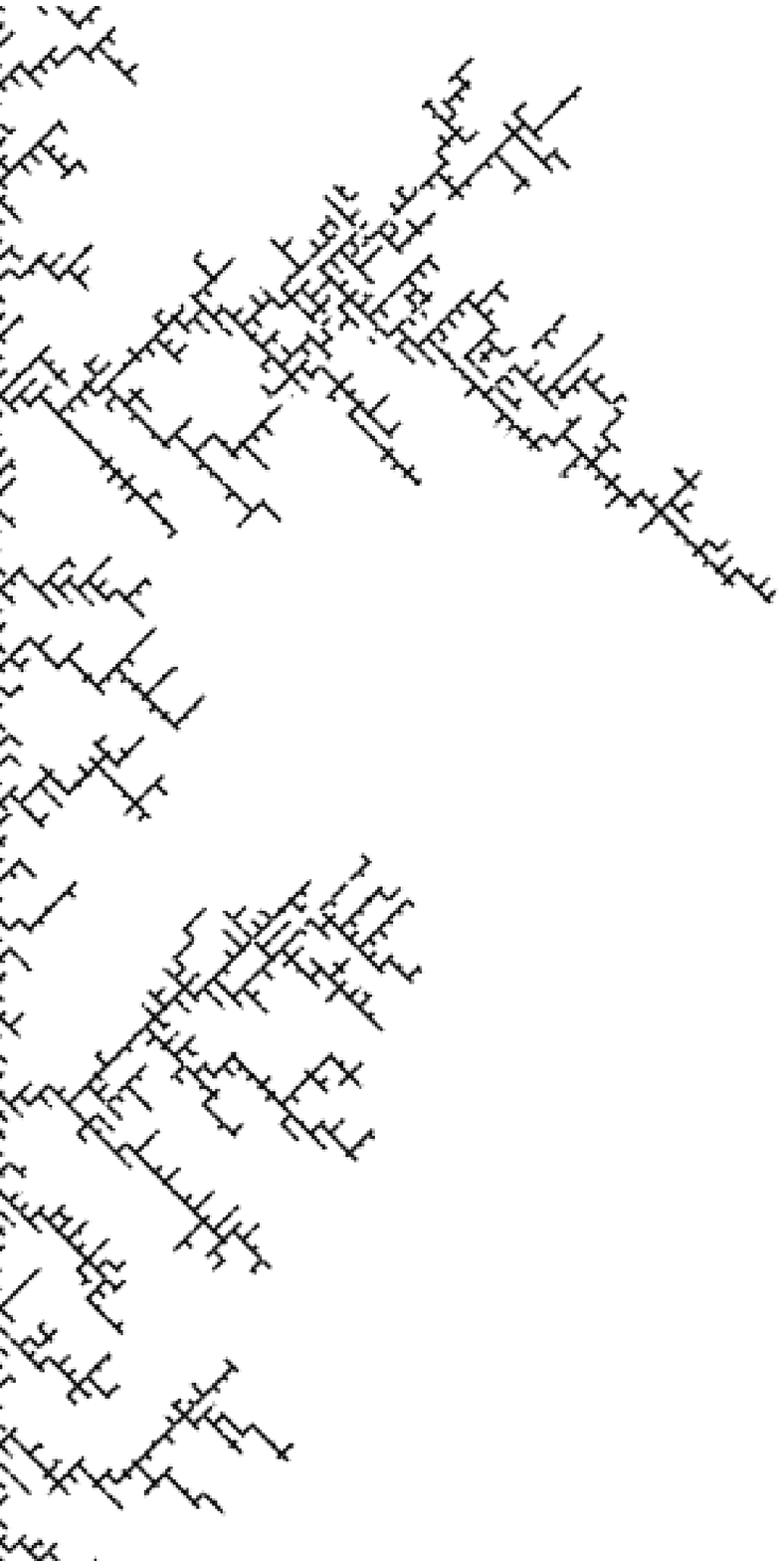}
\includegraphics*[scale=0.26]{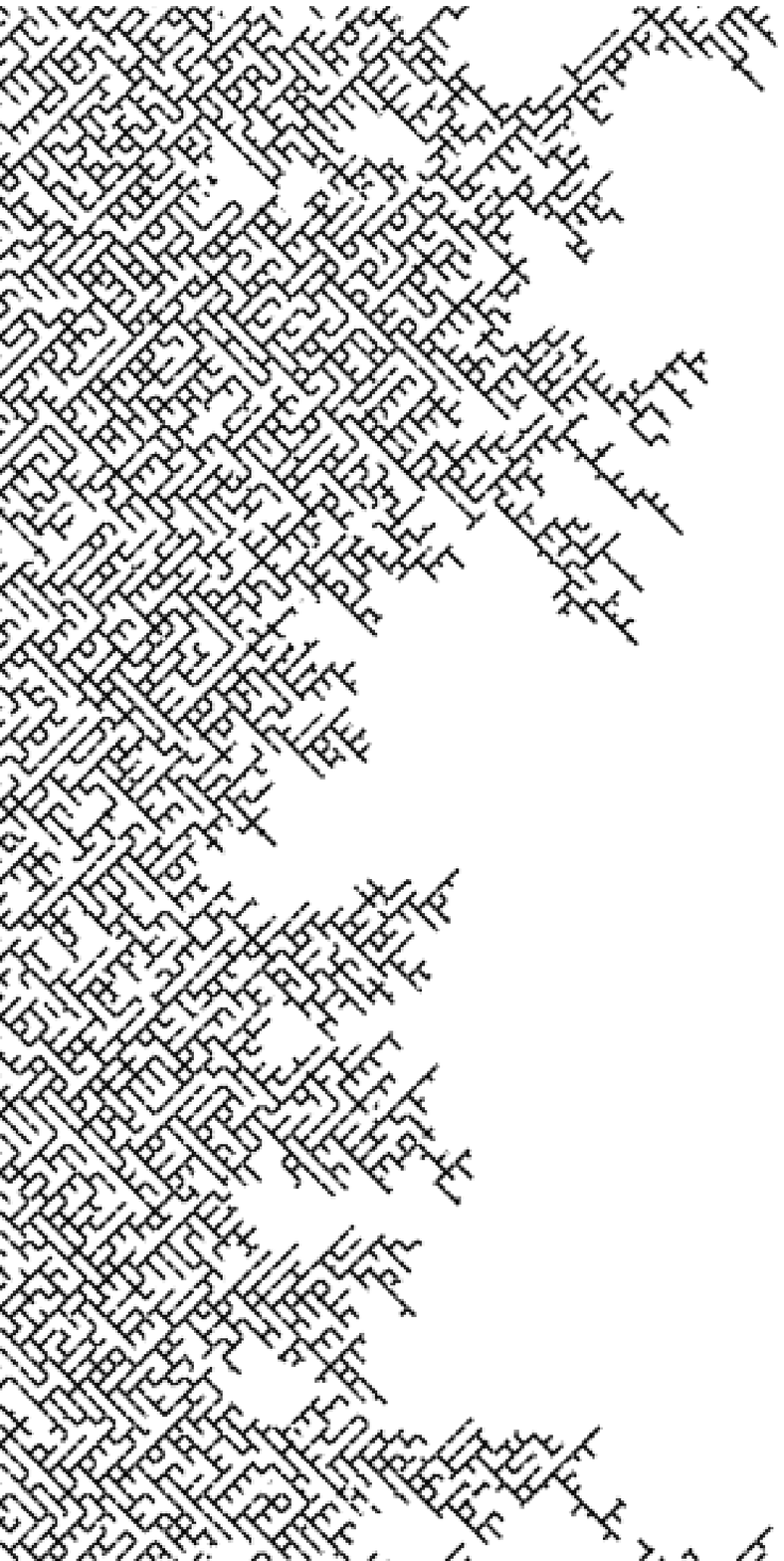}
\includegraphics*[scale=0.26]{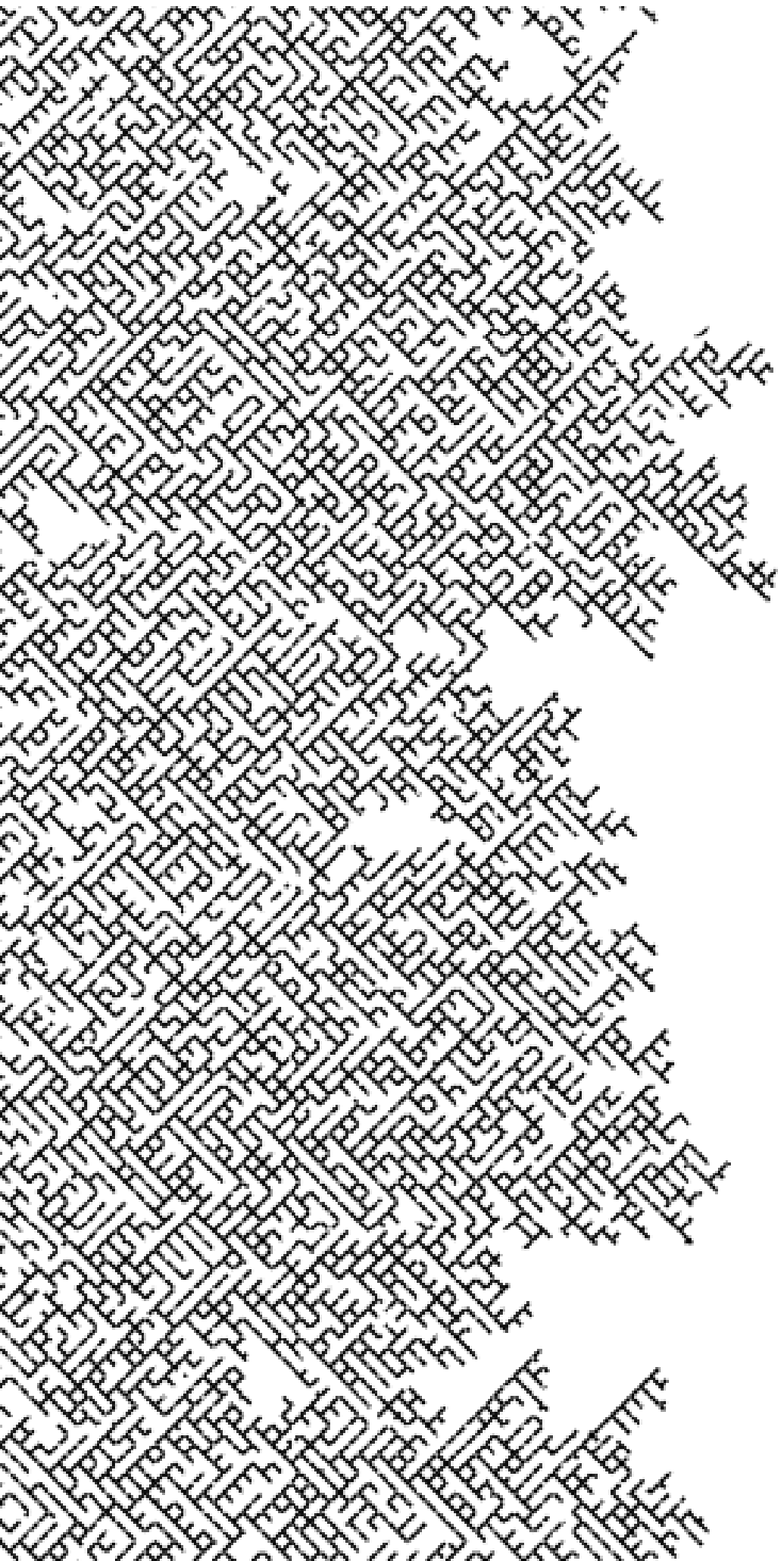}
\includegraphics*[scale=0.26]{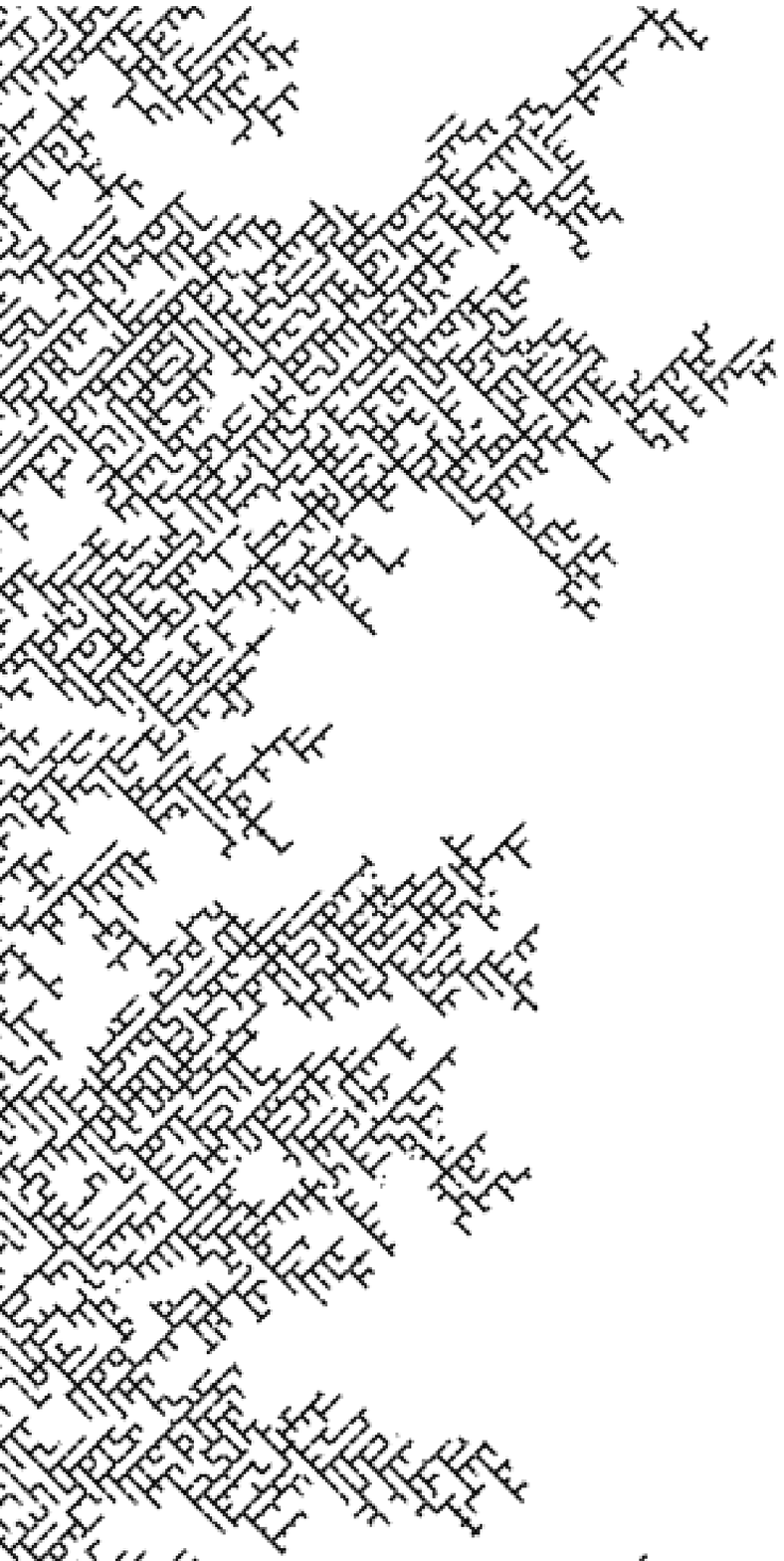}
\includegraphics*[scale=0.26]{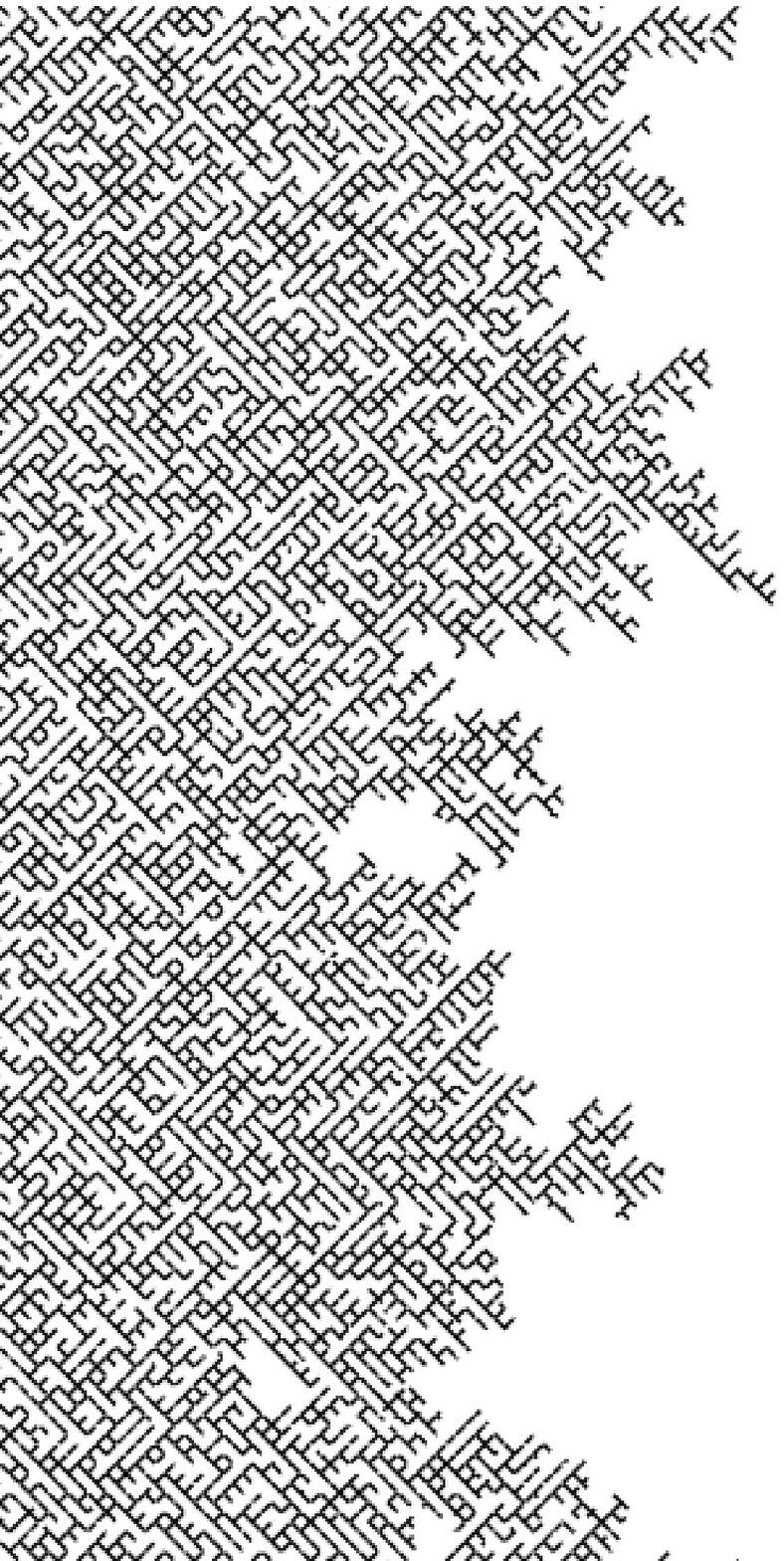}
%
%
\caption{Snapshots of the invading fluid near breakthrough for the
  finite viscosity ratio case, using distinct types of oil and
  temperature differences. All patterns have been generated with the
  same distribution of permeability for the random links and $L=80$.
  From top to bottom, $B$ changes from 7 to 3, while from left to
  right, $\Delta T$ assumes the values -4, 0 and 4.}
\label{fig:inv_V}
\end{center}
\end{figure}
Here we assume that the viscosity of the fluid phases typically obey
an exponential dependency on the inverse of temperature,
\begin{equation}
\mu_o = exp(B/T), \label{eq:viscosity}
\end{equation}
%
%
where the controlling physico-chemical parameter $B$, in units of
temperature, is considered to be constant. One calls a fluid ``heavy
oil'' if $B$ is high and ``light oil'' if $B$ is small.
%
%
In order to impose a temperature gradient across the medium, different
temperatures are assigned to the left (inlet) right (outlet)
boundaries of the lattice and we consider, then, a linear temperature
variation from inlet to outlet. This imposed gradient is constant in
time, which is analogous to assume that the thermal conductivities of
both fluids are negligible compared to that of the rock.
The temperature difference is defined as $\Delta T=T_{inlet}-T_{outlet}$,
and we carry out simulations many different dimensionless values of
$T_{inlet}$ and $T_{outlet}$ ($T_{inlet}$, $T_{outlet}$ = 1, 2, 3, 4
and 5) such that -4 $\leq$ $\Delta T$ $\leq$ 4.
%
%
Negative values of $\Delta T$ represent a cold injection which is not
of technological interest, but can help to understand the nature
beyond the fluid displacement under a temperature gradient.
In this model, both link length, $\ell$, and reference temperature,
$T_r$, are adjusting parameters which one can use to apply that model
to a practical application. For example, if one considers $T_r=
15^{\circ}°C$ and $\ell=1$cm (size of the fraction of the rock which
each link represent in the model), the temperature gradient varies
from 0 to 75 $^{\circ}°C/m$ as $\Delta T$ changes from 0 to 4.
%
%
In the next section, we show results for the two different cases of
the viscosity ratio. In both, $\mu_o$ is given by Eq.~(3)
%
%
while $\mu_I$ is constant, i.e., independent of the temperature. This
can be justified since, for example, water, which is a typical
invading fluid, has a viscosity that hardly changes, compared to many
types of oils, for common operational temperature intervals.
\begin{figure}
\begin{center}
\includegraphics*[scale=0.35]{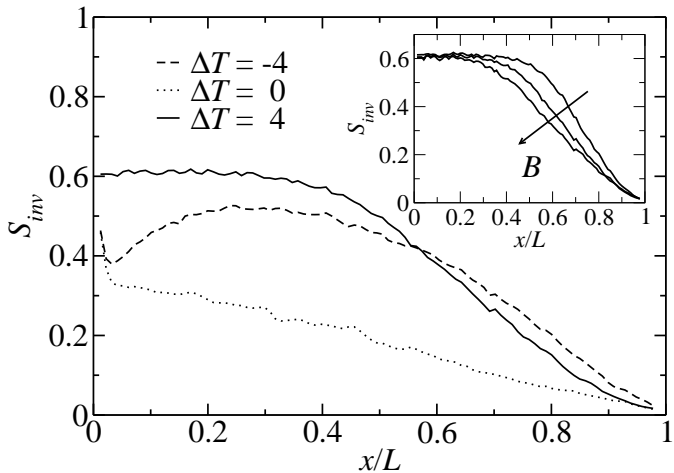}
\caption{Near-breakthrough saturation of the invading fluid for the
  finite viscosity ratio case. The curves correspond to three values
  of the temperature difference, $\Delta T=-4$, $0$, and $4$, $B=5$
  and $L=80$. In the {\it inset}, we also show the near-breakthrough
  saturation behavior, but for a fixed value of $\Delta T=4$, and
  $B=3$, $5$ and $7$.}
\label{fig:satura_V}
\end{center}
\end{figure}
%
%

\section{Results}

In this section, we study the patterns of the invaded region and the
percentage of recovered oil when an external fluid with unity
viscosity is injected into the medium.
%
%
According to our model, the viscosity ratio, $\mu_o/\mu_I$, has a
finite value which changes on $x$-direction, if $\Delta T$ is
different from zero. In Fig.~\ref{fig:inv_V} we show six different
patterns of the invading fluid with the same distribution of random
links for different values of $\Delta T$ and $B$. From top to bottom
we change $B$ from $7$ to $3$, while from left to right $\Delta T$
assumes the corresponding values of $-4$, $0$ and $4$. We clearly
observe that either decreasing $B$ or for positive values of $\Delta
T$, the invading patterns become more compact.

In the isothermal case, $\Delta T=0$, the oil viscosity is a constant
depending only on $B$. For a heavy oil ($B=7$), we can see finger
patterns appearing in the invaded region. When $\Delta T>0$, the
viscosity ratio is small on the left side of the lattice and becomes
larger as the invading fluid penetrates in $x$-direction.  First, the
front advances compactly but at some time, a finger appears and grows
faster until it reaches the other end of the lattice. In the opposite
case, when $\Delta T$ $<$ 0, fingers appear initially and then become
broader. For $B=3$, the changes in the viscosity ratio due to the
temperature difference suppress the appearance of fingers.
\begin{figure}
\begin{center}
\includegraphics*[scale=0.35]{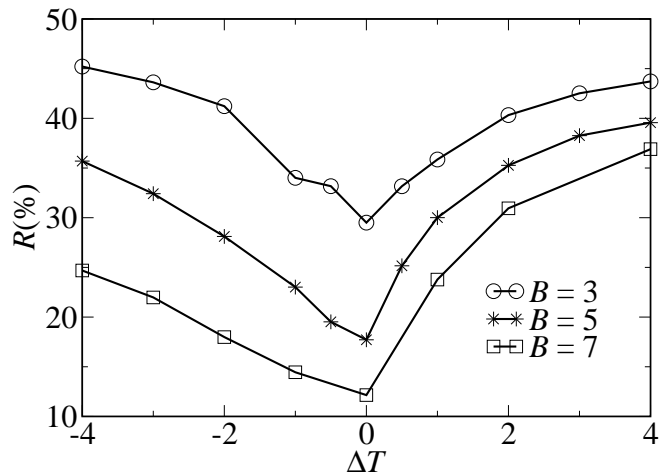}
\caption{Percentage of recovered oil versus $\Delta T$ for the finite
  viscosity ratio case using three different values of $B$ and $L=80$.}
\label{fig:recovery_V}
\end{center}
\end{figure}

For each set of parameters, we perform simulations with $50$
realizations of the disordered porous media, to obtain average values
of both the invading fluid saturation and the percentage of recovered
oil at the breakthrough. In Fig.~\ref{fig:satura_V}, we show the
near-breakthrough saturation profile of the invading fluid for
different values of $\Delta T$ and $B$. We observe that for a positive
$\Delta T$ the saturation profile has a plateau before it starts
decreasing in the $x$-direction. However, for a negative $\Delta T$,
finger patterns in the beginning of the lattice produce a dip in the
saturation profile close to the inlet. The percentage of recovered oil
versus $\Delta T$ is shown in Fig.~\ref{fig:recovery_V} for three
different values of $B$. We see that, despite heavy oil recovery is
smaller, all types of oil tend to the recovery performance for high
values of $\Delta T$.

Now, we study the recovery of an oil that is much heavier than the
invading fluid, i.e., for the case of an extremely large viscosity
ratio, $\mu_o/\mu_I \rightarrow \infty$. This idealized condition is
implemented here by considering that the pressure in the invaded fluid
immediately adjusts to the injection pressure, i.e., the invading
fluid is inviscid.
%
%
%
This simplification allows us to simulate bigger lattice sizes, since
we need to solve Eqs.~(2) only for non-invaded sites.
\begin{figure}
\begin{center}
\includegraphics*[scale=0.31]{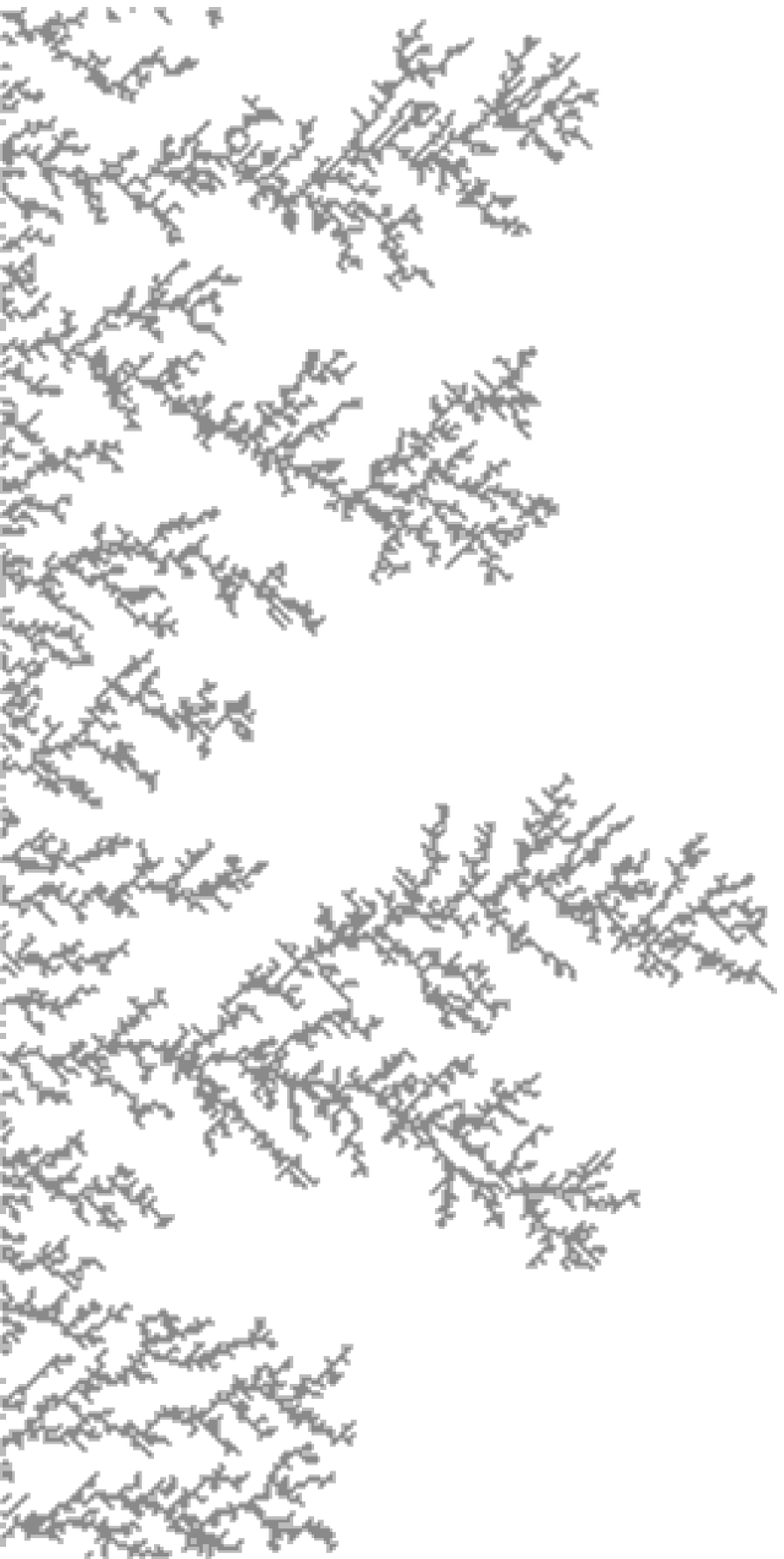}
\includegraphics*[scale=0.31]{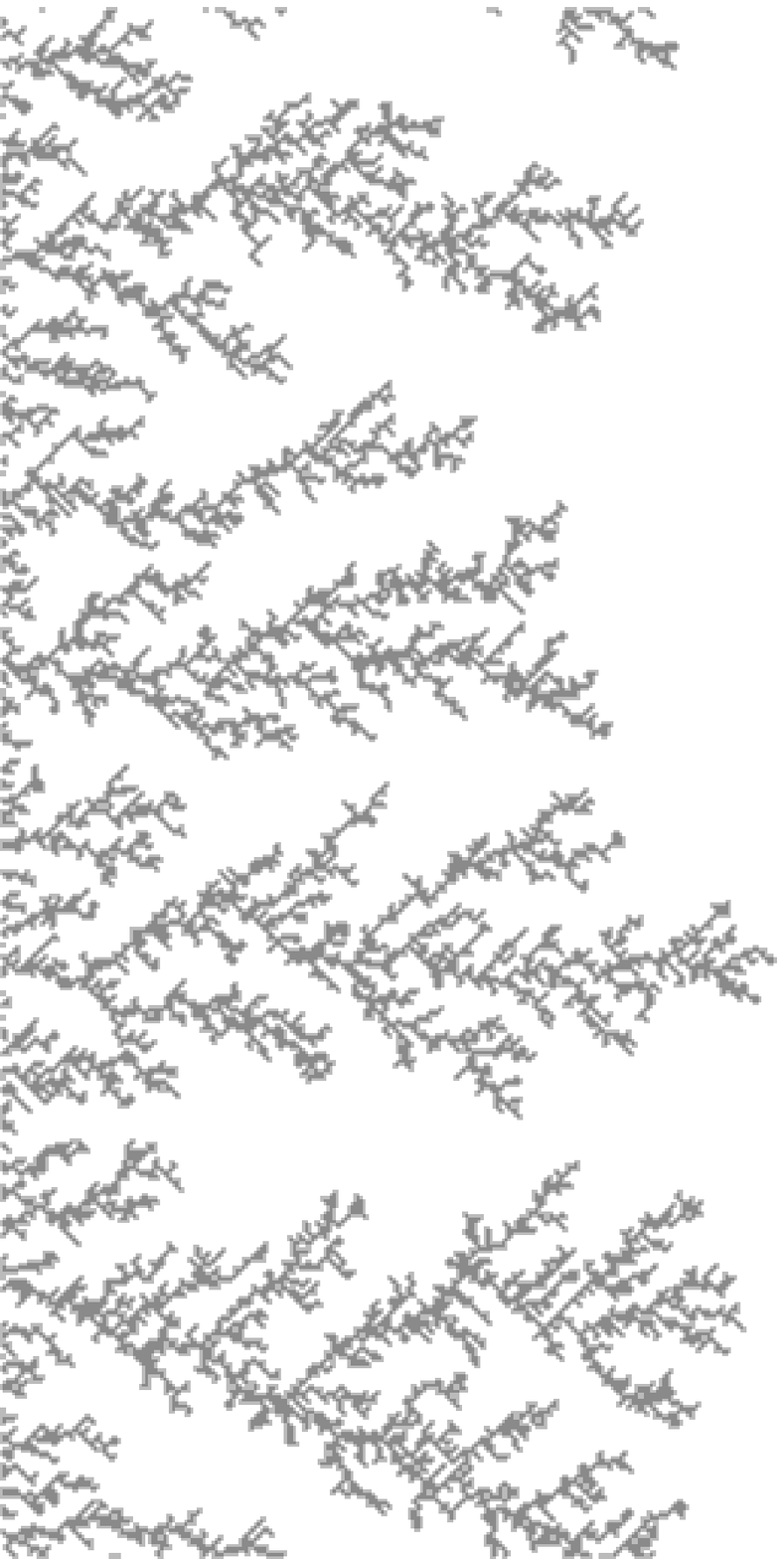}
\includegraphics*[scale=0.31]{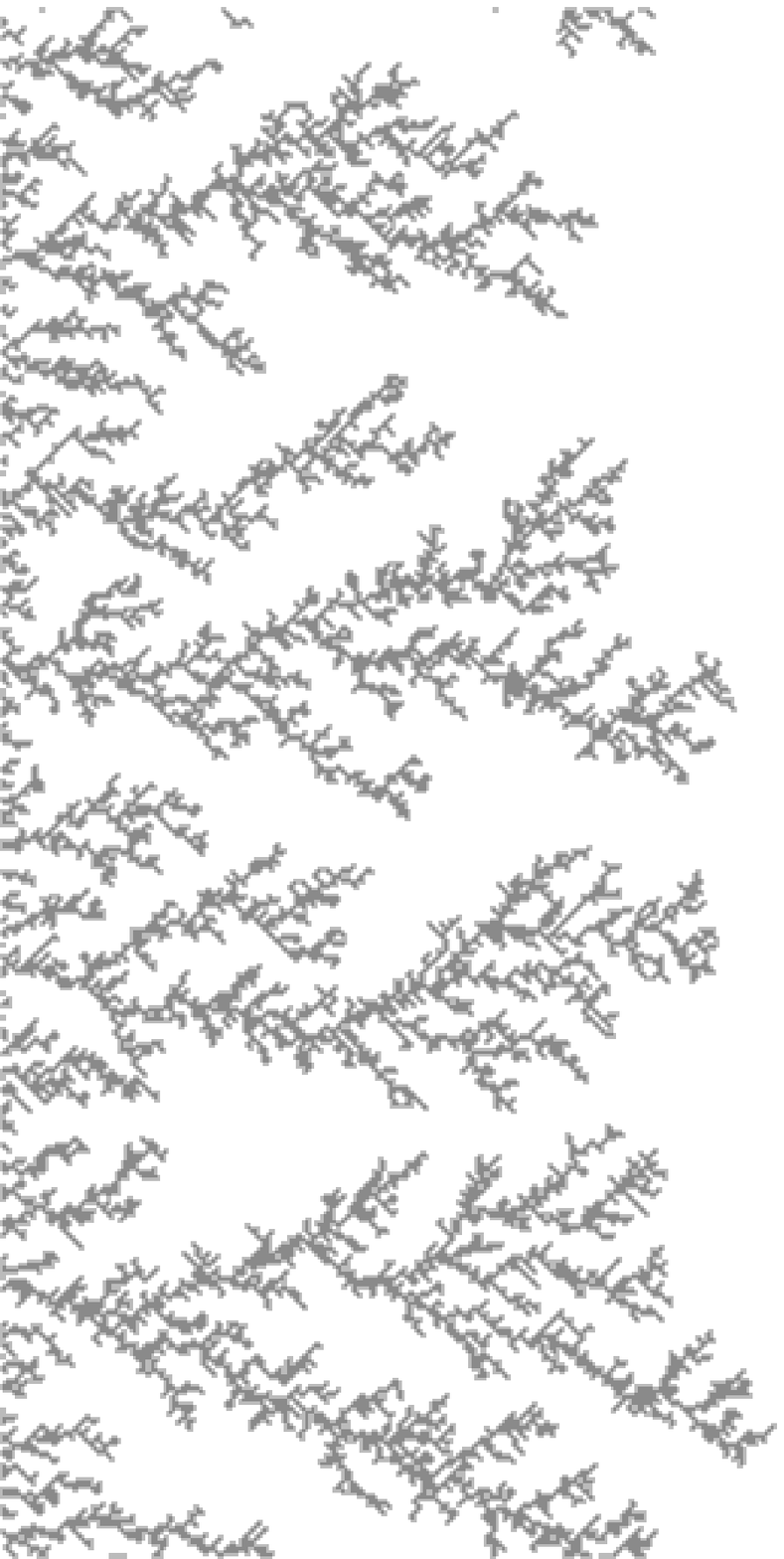}
\caption{Snapshots of the invading fluid near breakthrough for the
  infinite viscosity ratio case. All patterns have been generated with
  the same distribution of permeability for the random links, $B=5$
  and $L=256$. From left to right, $\Delta T=-4$, $0$ and $4$.}
\label{fig:inv}
\end{center}
\end{figure}

In this case, the observed patterns of the interface are always
viscous fingering-like, as shown in Fig.~\ref{fig:inv}, for $\Delta
T=-4$, $0$ and $4$, with $B=5$ and $L=256$. For $\Delta T=0$, these
patterns show fingers which agree with well know two-phase
displacement patterns with infinite viscosity ratio
\cite{homsy,chen}.
%
%
We can also see that different patterns occur for different values of
$\Delta T$.  The reason for this behavior is that, despite the
inviscid characteristic of the defending fluid, the oil viscosity has a
finite value given by Eq.~(3) which falls as the temperature is raised.
Then the number of longer fingers for $B=5$ increases with $\Delta T$
as shown Fig.~\ref{fig:inv}. This observation is also valid for $B=1$
and $3$ (not shown), and therefore represents a standard behavior in
the case of infinite viscosity ratio.
\begin{figure}
\begin{center}
\includegraphics*[scale=0.35]{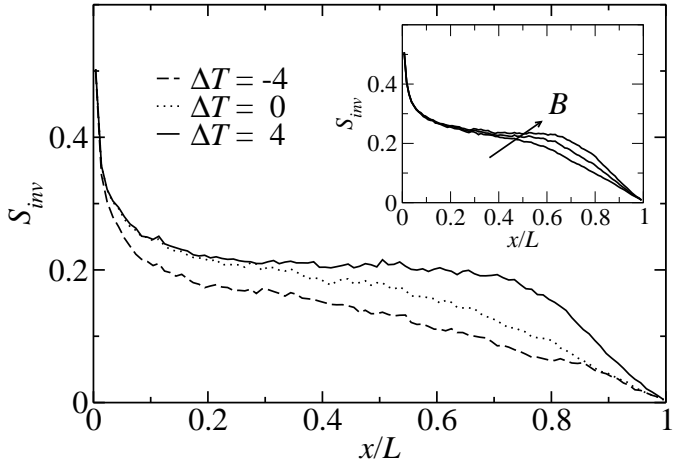}
\caption{Near-breakthrough saturation of the invading fluid for the
  infinite viscosity ratio case. The curves correspond to three values
  of the temperature difference, $\Delta T=-4$, $0$, and $4$, $B=5$
  and $L=256$. In the {\it inset}, we also show the near-breakthrough
  saturation behavior, but for a fixed value of $\Delta T=4$, and
  $B=1$, $3$ and $5$.}
\label{fig:satura}
\end{center}
\end{figure}

For each set of parameters, we averaged over 100 realizations to
obtain the invading fluid saturation and the percentage of recovered
oil. The near-breakthrough saturation profile of the invading fluid is
shown in Fig.~\ref{fig:satura} for different values of the relevant
parameters. For a negative value of $\Delta T$, the saturation profile
always decays in $x$-direction. However, for a positive $\Delta T$, we
can identify a region in the center of the medium with approximately
constant saturation. The extension of this region increases with $B$
because heavier oils create slower and wider fingers, which tend to
make the invasion patterns more compact.
\begin{figure}
\begin{center}
\includegraphics*[scale=0.35]{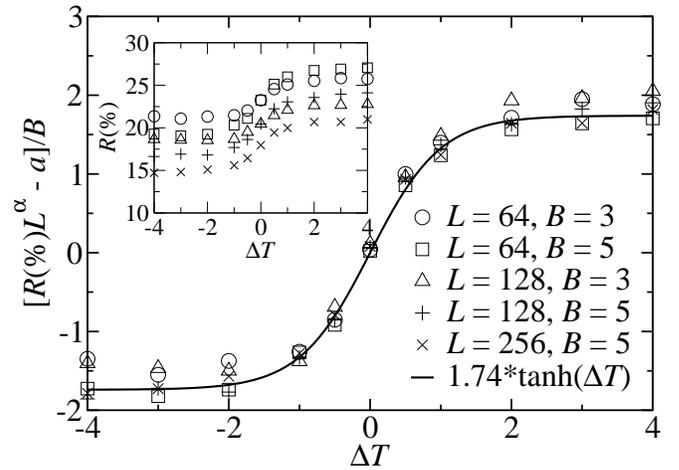}
\caption{Percentage of recovered oil versus $\Delta T$ calculated in
  the infinite viscosity ratio case, for different values of $L$ and
  $B$. In the main plot, after proper rescaling, we show that these
  curves collapse on the top of each other and can be all fitted to a
  hyperbolic tangent function. In the {\it inset}, we show the same
  results before rescaling.}
\label{fig:recovery}
\end{center}
\end{figure}

In the inset of Fig.~\ref{fig:recovery}, the percentage of recovered
oil $R(\%)$ versus $\Delta T$ is shown for different values of $L$ and
$B$. The main plot shows that, after rescaling, these curves collapse
on the top of each other and can be closely described by a hyperbolic
tangent function in the form,
\begin{equation}
  R(\%) = \frac{a+bB \tanh(\Delta T)}{L^\alpha},
\label{eq:equation}
\end{equation}
%
%
with parameters $a=51.11$ and $b=1.74$ obtained through the best
nonlinear fit to the data. The exponent $\alpha$ is found to be about
$0.2$ and can be computed as $\alpha=d-d_f$, where $d=2$ is the
Euclidian dimension of the lattice and $d_f$ is the fractal dimension,
which is found to be $1.8$ for our system. The results show that above
a certain value of temperature difference, $\Delta T \approx 2$, no
relevant change in $R(\%)$ is observable anymore for all types of oil.
This means that a too strong temperature gradient can be an
unnecessary cost to the recovery process.
\begin{figure}
\begin{center}
\includegraphics*[scale=0.35]{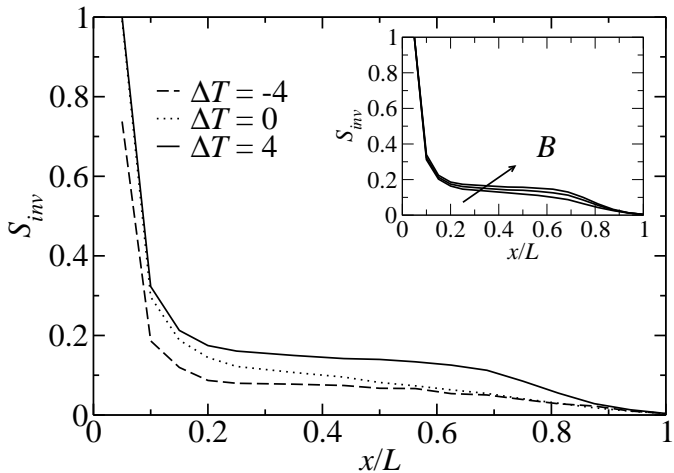}
\caption{Near-breakthrough saturation profile of the invading fluid
  penetrating in a cubic lattice for the case of infinite viscosity
  ratio.  The curves are the results of simulations for $\Delta T=-4$,
  $0$, and $4$, with parameter $B=5$, and system size $L=20$. In the
  {\it inset}, we also show the near-breakthrough saturation behavior,
  but for a fixed value of $\Delta T=4$, and $B=3$, $5$ and $7$.}
\label{fig:satura3D}
\end{center}
\end{figure}

\begin{figure}
\begin{center}
\includegraphics*[scale=0.35]{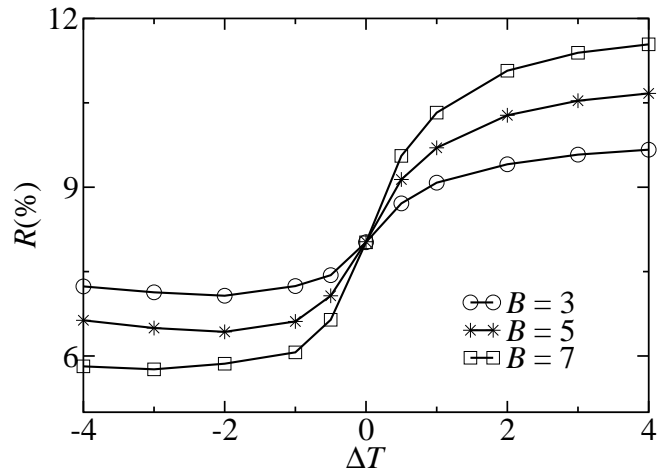}
\caption{Dependence of the percentage of recovered oil on the
temperature difference $\Delta T$ or different values of $B$ on a
cubic lattice with $L$ = 20 for infinite viscosity
ratio.}\label{fig:recovery3D}
\end{center}
\end{figure}

Finally, we also performed simulations with our model in the infinite
viscosity ratio regime using a more realistic three-dimensional porous
medium substrate. More precisely, we obtained results for a cubic
lattice with size $L=20$ and averaged over 100 realizations for
several values of $B$ and $\Delta T$. As shown in
Figs.~\ref{fig:satura3D} and \ref{fig:recovery3D}, the saturation
profiles and the recovery performance of the system, respectively,
remain basically the same as compared to the results found for
two-dimensional porous media models (for comparison, see Figures
\ref{fig:satura} and \ref{fig:recovery}). This qualitative similarity
reinforces the validity of our approach as a way to increase the
efficiency of the recovery process by means of a temperature gradient.
%
%
\section{Conclusions}

In summary, the main purpose here was to investigate how the front
between two immiscible fluids propagates in a model porous medium as a
function of the ratio of their viscosity and under the influence of an
imposed global temperature gradient. An obvious technological
application of our study would be to enhance the recovery efficiency
of oil being pushed by hot-water in a petroleum reservoir.  In our
simulations, this has been accomplished by explicitly coupling the oil
viscosity with the inverse of temperature locally in terms of a simple
exponential dependence. We thus proceeded with the displacement of
several types of oil through many realizations of disordered porous
medium, and subjected to a range of distinct temperature differences.

Two different regimes of viscosity ratio have been studied. In the
first, oil viscosity changes with temperature and the viscosity ratio
is always finite, but can vary over several orders of magnitude. In
the second regime, oil is assumed to be ``heavy'', in the sense that
it is a extremely viscous when compared do the invading fluid, even if
a maximum temperature difference is applied, hence the viscosity ratio
can be considered as infinite. We find that the best conditions for
recovery are significantly dependent on the adopted regime. In the
finite viscosity ratio case, an oil with a viscosity that is only
weakly dependent on the temperature is better recovered, independently
on the strength or sign of the gradient. Also in this case, different
invasion patterns can be observed as the viscosity ratio changes,
namely, we find fronts that are compact, unstable and sometimes a
mixing of both.

Differently, in the infinite viscosity ratio case, oil recovery
increases with the exponential parameter $B$ if the temperature
difference is positive.  Moreover, recovery is found to follow a
hyperbolic tangent behaviour on the temperature difference and the
best recovery is obtained for positive temperature gradients
(Hot-Water Injection). It would be interesting to verify our
predictions with experimental results. Since field experiments in this
area are usually difficult and expensive, it would be advisable to run
laboratory-size experiments in which interface patterns are
dynamically registered and the recovered volume is systematically
measured while a viscous fluid is displaced by a less viscous one in a
porous medium under the influence of temperature gradients.

\begin{acknowledgments}
  We thank the Brazilian Agencies CNPq, CAPES, FUNCAP and FINEP, the
  FUNCAP/CNPq Pronex grant, Petrobras and the National Institute of
  Science and Technology for Complex Systems for financial support.
\end{acknowledgments}

\end{document}